\documentclass[twocolumn,prb,aps,floatfix]{revtex4}
\usepackage{graphicx}
\usepackage{dcolumn}
\usepackage{bm}
\begin{document}

\newcommand{\nm}{
\ensuremath{\rm\,nm}}
\newcommand{\mph}{
\ensuremath{\rm\,\mu m/hour}}
\newcommand{\um}{
\ensuremath{\rm\,\mu m}}
\newcommand{\V}{
\ensuremath{\rm\,V}}
\newcommand{\dBm}{
\ensuremath{\rm\,dBm}}

\author{J.~R.~Gell\footnote{also at Toshiba Research Europe Limited, Cambridge Research Laboratory, 260 Cambridge Science Park, Cambridge, CB4 0WE, United Kingdom. electronic mail: jrg31@cam.ac.uk}}
\author{P.~Atkinson}
\author{S.~P.~Bremner}
\author{F.~Sfigakis}
\author{M.~Kataoka}
\author{D.~Anderson}
\author{\\G.~A.~C.~Jones}
\author{C.~H.~W.~Barnes}
\author{D.~A.~Ritchie}
\affiliation{Cavendish Laboratory, University of Cambridge, J.~J.~Thomson Avenue, Cambridge, CB3 0HE, United Kingdom}

\title{Surface-acoustic-wave-driven luminescence from a lateral p-n junction}
\author{M.~B.~Ward}
\author{C.~E.~Norman}
\author{A.~J.~Shields}
\affiliation{Toshiba Research Europe Limited, Cambridge Research Laboratory, 260 Cambridge Science Park, Cambridge, CB4 0WE, United Kingdom}

\begin{abstract}
The authors report surface-acoustic-wave-driven luminescence from a
lateral p-n junction formed by molecular beam epitaxy regrowth of a
modulation doped GaAs/AlGaAs quantum well on a patterned GaAs
substrate. Surface acoustic wave driven transport is demonstrated by
peaks in the electrical current and light emission from the GaAs
quantum well at the resonant frequency of the transducer. This type
of junction offers high carrier mobility and scalability. The
demonstration of surface acoustic wave luminescence is a significant
step towards single-photon applications in quantum computation and
quantum cryptography.
\end{abstract}

\maketitle

Surface acoustic waves (SAWs) travelling on a piezoelectric
material containing a buried layer of charge have attracted
significant interest in recent years.\cite{Talyanskii1997,
Stotz2005, Wixforth1986} In particular it has been shown that the associated
electrostatic wave can be used to transport single electrons
through a split gate leading to a quantised
current.\cite{Shilton1996} This effect led to the proposal of a
single-photon source in which the SAW is used to pump single
electrons across a lateral p-n junction.\cite{Foden2000} In this proposal the
photon emission time coincides with the arrival of
each SAW minimum, creating a high repetition rate single-photon
source ideal for use in quantum cryptography. Barnes \textit{et
al}. have proposed a computation scheme where information is
stored in the spin of an electron propagating in a SAW
minimum.\cite{Barnes2000} Single-qubit and two-qubit operations would be
performed as the electrons are carried through a series of
magnetic and non-magnetic gates. Our device could act as an
optical readout for this scheme with quantum information transferred
from the electron spin to the polarization of the emitted
photons.\cite{Vrijen2003}

\begin{figure}
\centering
\includegraphics[width=250px, height=250px]{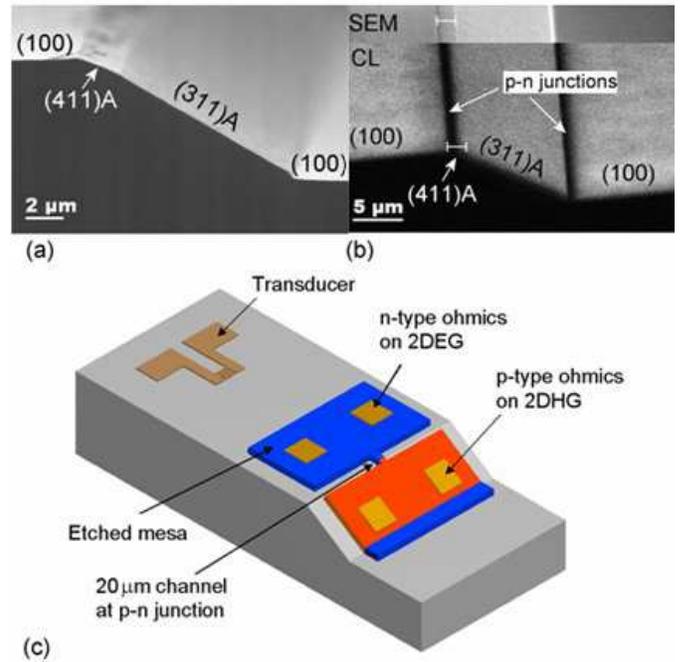}
\caption{(a) Scanning electron microscope image (SEM) showing a
cross-section through the wafer after re-growth.  p-n junctions
form at the interfaces between the planes.  (b) Panchromatic
cathodoluminescence  (CL) image at room temperature taken at an
oblique angle. Inset: SEM image for comparison. (c) Schematic
diagram of the device. } \label{Fig1}
\end{figure}

The first step towards such a single-photon source is the
demonstration of SAW-mediated transport across a lateral p-n
junction and the resulting luminescence.  Several types of lateral
junctions have been investigated so far.\cite{Hosey2003,
Kaestner2002, Nash2005, Cecchini2004} In this letter we present
results from a lateral p-n junction in a GaAs/AlGaAs modulation
doped quantum well formed by molecular beam epitaxy regrowth on a
patterned GaAs wafer.\cite{Vaccaro1998, North1999, Ocampo2003} The
amphoteric nature of silicon enables n-doped material to be grown on
the flat (100) GaAs substrate and p-doped material to be grown on
etched (n11)A facets (for n$\leq$3) in the same growth
step.\cite{Wang1985} A lateral p-n junction is formed at the
interface between the flat and etched planes. Prior to growth the
(100) GaAs substrate was patterned with stripes parallel to the
[1\={1}0] direction using 18:11:180 BHF:H$_{2}$O$_{2}$:H$_{2}$O
solution at $10^{\circ}$C. This etch exposes $\sim25^{\circ}$ facets
to the (100) surface, as shown in Figure \ref{Fig1}a, creating a
(100)-(311)\,A-(100) step. The photoresist is removed by a series of
solvent rinses and an oxygen plasma etch before the wafer is
transferred to the MBE system and cleaned in-situ by exposure to a
hydrogen radical beam.\cite{Burke1999} The growth consisted of a
$5.5 \times 10^{11}\,\mathrm{cm^{-2}}$ Si delta doped layer to
counteract the effects of residual surface
contamination,\cite{Burke1999} a $100\nm$ Al$_{0.33}$Ga$_{0.66}$As
buffer layer, a 500$\nm$ (2.5$\nm$ Al$_{0.33}$Ga$_{0.66}$As/
2.5$\nm$ GaAs) superlattice barrier, a 15$\nm$ GaAs quantum well, a
40$\nm$ Al$_{0.33}$Ga$_{0.66}$As spacer, a 40$\nm$ $1.1 \times
10^{18}\,\mathrm{cm^{-3}}$ Si doped Al$_{0.33}$Ga$_{0.66}$As region
and a $14\nm$ GaAs cap. The GaAs growth rate was $1\mph$, the
substrate temperature was $600^{\circ}$C and the V/III beam
equivalent pressure ratio was 7.4. This forms a 2D lateral n-p-n
junction with a high mobility two-dimensional electron gas (2DEG) on
the top plane (carrier density of $3.4 \times
10^{11}\,\mathrm{cm^{-2}}$ and mobility of $4.9 \times
10^{5}\,\mathrm{cm^2/Vs}$ measured at $1.5\,$K after illumination
with a red light emitting diode) and a two-dimensional hole gas
(2DHG) on the (311)\,A facet. Figure \ref{Fig1}a also shows an
additional (411)\,A facet formed at the top (100)-(311)\,A junction
during the overgrowth. Figure \ref{Fig1}b shows a room temperature
panchromatic cathodoluminescence image of the structure. The
positions of the p-n junctions at the top and bottom of the facet
show up as dark lines. In these regions the high electric field
sweeps the carriers away before they are able to recombine. The
width of these lines can be used to estimate the depletion regions
of the unbiased junction as $\sim$$\,1\um$.

Devices were fabricated on this wafer as shown in Figure
\ref{Fig1}c. Although the wafer has a p-n junction at both the top
and bottom of the facet all measurements were carried out on the
top junction as this allowed better lithographic resolution.
Standard optical lithography was used to pattern a mesa giving a
$20\um$ wide channel across the top p-n junction. AuBe p-type
contacts were made to the 2DHG and AuGeNi n-type contacts to the
top 2DEG.  An interdigital transducer with a resonant frequency of
approximately 984\,MHz was placed $2.2\,\mathrm{mm}$ away from the
junction on an etched region of the top plane. The device was
bonded into a package designed to provide some shielding from the
free-space electromagnetic (EM) wave radiating from the transducer
circuit. Measurements were carried out in a liquid helium
continuous flow cryostat with a base temperature of $4\,$K. The
actual device temperature is not well known, as the microwave
coaxial cables used to apply radio-frequency (RF) signals to the
transducer will cause some heating.  The light emitted was
collected through a microscope objective lens and spectrally
recorded with a grating spectrometer and nitrogen cooled charge
coupled device (CCD).

\begin{figure}
\includegraphics[scale=1.2]{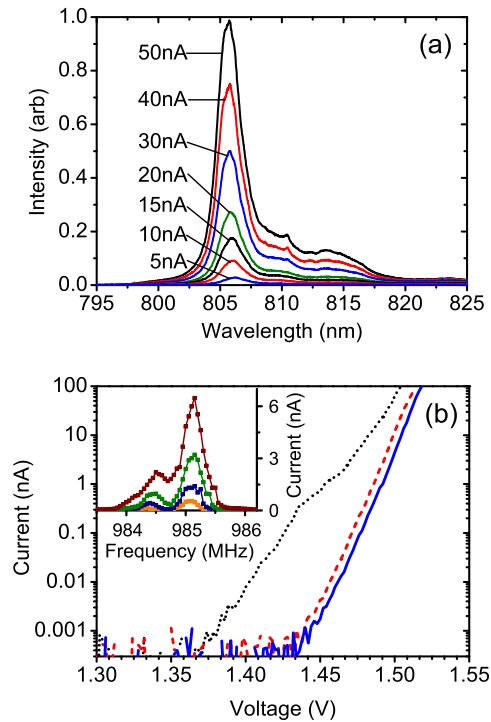}
\caption{(a) Electroluminescence spectra for a range of bias
current ($5$--$50\,{\rm nA}$) with no SAW present. (b)
IV-characteristics of the p-n junction.  Solid line (blue): no
power applied to the transducer, dotted line (black): $12\dBm$
applied at the transducer's resonant frequency, dashed line (red):
$12\dBm$ applied off-resonance.  The inset shows the frequency
dependence of the current for a range of bias voltages
($1.42$--$1.45\V$).} \label{Fig2}
\end{figure}

The electrical characteristics of several diodes were studied
without the presence of a SAW.  They showed the expected
rectifying behavior with negligible reverse-bias current ($<$$\,10\,\rm {pA}$ in the
range 0 to -2\,V) and at a fixed forward bias voltage
the current through the device was stable.

Figure \ref{Fig2}a shows the emission from one device for a range of
forward bias conditions.  The peak at $806\nm$ and the broad peak at
longer wavelength are attributed to emission from the GaAs quantum
well. The quantum well is nominally $15\nm$ wide but there will be
some lateral variation in the width due to the differences in the
growth rates on different crystallographic planes. In particular the
quantum well should be narrower on the angled (311)\,A facets since
the incident Ga flux is about 90\% of that on the perpendicular
(100)\,planes. The Ga adatom diffusion length is also longer on the
(311)\,A facets than on the (100)\,planes, leading to the buildup of
material and the formation of extra (411)\,A and (111)\,A facets at
the top and bottom interfaces
respectively.\cite{Hata1990,Takebe1997} Both of these effects will
cause the quantum well width to vary across the p-n junction, so the
wavelength of the emitted photons will depend on the exact position
of the recombination.  Emission was seen at the top of the facet
across the width of the $20\um$ channel, but the intensity and exact
form of the spectra varied across the channel. This is not
unexpected as there are some irregularities in the facet caused by
unevenness in the pre-regrowth etch and there are likely to be
associated fluctuations in the well width. The position of the peak
at $806\nm$ is seen to decrease slightly in wavelength at higher
bias voltages.

With no SAW present the diode characteristic of this device was
exponential (Figure \ref{Fig2}b solid line).  However, the presence
of the SAW changed the IV-curve dramatically (Figure \ref{Fig2}b
dotted line). This effect was not seen when the transducer was
driven off-resonance (Figure \ref{Fig2}b dashed line) indicating
that it is due to SAW-mediated transport across the junction. There
is a small increase in the current off-resonance, which might be
caused by heating of the diode or modulation of the junction by the
free-space EM wave but it is insignificant compared to the effect of
the SAW generated on-resonance. The inset in Figure 2b shows the
change in the current as a function of transducer frequency. The
oscillations are caused by interference between the main SAW and a
reflected SAW or the free-space EM wave.\cite{Astley2006} It can be
seen from Figure \ref{Fig2}b that below a certain bias voltage there
is no SAW-driven current. This is expected as the electric field of
the unbiased junction ($1.5\,$V dropped across the depletion width
of $\sim$$\,1\um$) is too large for the SAW (wavelength $3\,\um$,
amplitude of a few 10s of $mV$) to transport electrons across the
junction.\cite{Robinson2001,Schneble2006} Subsequent measurements
were taken when the diode was forward biased to below threshold
(where negligible light emission was observed) but the potential
slope was shallow enough for the SAW to carry electrons across the
junction. Over time the resonant frequency of the transducer
decreased approximately linearly.  It is thought that this is due to
mass loading of the transducer as residual gases in the cryostat
condensed on the cold sample. \cite{Campbell1989} Although this
shift is not important it means that the exact frequency dependence
of measurements taken at different times cannot be compared.

\begin{figure}
\includegraphics[scale=1.5]{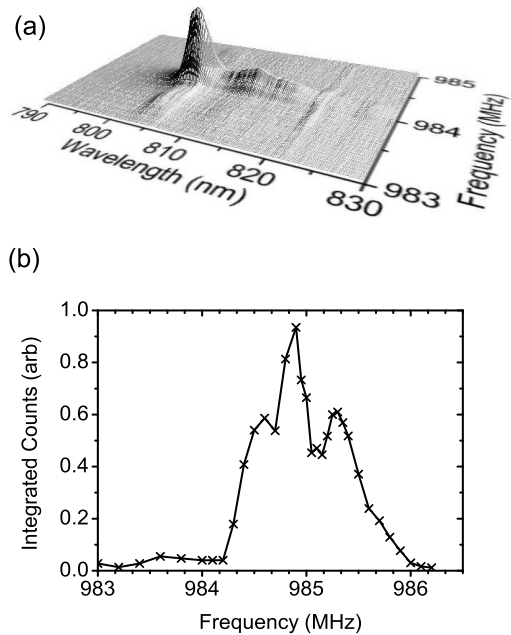}
\caption{(a) SAW-driven luminescence as a function of the frequency
applied to the transducer ($12\dBm$).  The diode was d.c.\ biased to
$1.47\V$. (b) Integrated light intensity at $1.47\V$, with a
transducer power of $12\dBm$ as a function of the frequency applied
to the transducer. } \label{Fig3}
\end{figure}

Figure \ref{Fig3}a shows the luminescence from the junction as a
function of the frequency applied to the transducer.  The peak in
the emission occurs at the resonant frequency of the transducer
where a SAW is generated.  This associates the luminescence with the
SAW as the pick-up from the free-space EM wave is not strongly
frequency dependent.\cite{Gell2006}  Away from the resonant
frequency of the transducer there is no light emission observed
above the background.  The luminescence is dominated by emission
from the quantum well showing that the carriers are confined in the
well across the whole width of the junction and that there are no
other radiative paths.  The peak emission occurs at
$\sim$$\,805.5\nm$, slightly lower than seen in the
electroluminescence measurements. The majority of the SAW-driven
light was localized across a small region  corresponding to a
visible irregularity in the facet caused by the pre-regrowth etch.
Figure \ref{Fig3}b shows the integrated intensity of the light
emitted as a function of transducer frequency.  The oscillations are
due to the same interference effect as observed in the current
traces.

In summary, we have fabricated a high mobility lateral diode by MBE
regrowth and shown emission from the $15\nm$ quantum well.
SAW-driven electron transport across the junction has been
demonstrated by an increase in both the current through and
luminescence from the junction at the resonant frequency of the
transducer. This device could be used for optical readout or
information transfer in a quantum computation scheme and is the
first step towards realizing an acousto-electrically driven
single-photon source for high frequency operation.

The auothrs would like to thank K.~Cooper for help and advice on
device processing. This research is partly funded in QIP IRC
(GR/S82176/01).  One of the authors (J.~R.~G) would like to thank
EPSRC and TREL for funding.


\begin{references}
\bibitem{Gell2006} This is confirmed by additional pulsed measurements.
\end{references}
\end{document}